\documentclass[a4paper,12pt]{article}

\usepackage{caption}
\usepackage{subcaption}

\usepackage[english]{babel}
\usepackage{fancyhdr}
\usepackage{setspace}
\usepackage{graphics}
\usepackage{graphicx}
\usepackage{eso-pic}
\usepackage{amsmath}
\usepackage{amstext}
\usepackage{amsfonts}
\usepackage{amssymb}
\usepackage{hyperref}
\usepackage{slashed}
\usepackage{cite}
\usepackage{bbold}
\usepackage{multirow}
\usepackage{esdiff}
\usepackage[font=small,labelfont=small]{caption}
\usepackage{commath}

\usepackage{mathabx}
\usepackage{relsize}

\usepackage{enumitem}

\usepackage{tabularx}

\usepackage{mathrsfs} 

\usepackage[affil-it]{authblk} 
 
\renewcommand{\vec}[1]{\ensuremath{\boldsymbol{#1}}}

\newcommand{\R}{\mathbb{R}}	
\newcommand{\iu}{\mathrm{i}}	
\newcommand{\ee}{\mathrm{e}}	
\newcommand{\F}{\mathcal{F}} 
\newcommand{\Scri}{\mathscr{I}}

\newcommand{\bra}[1]{\ensuremath{\left< #1\,\right|}}
\newcommand{\ket}[1]{\ensuremath{\left|\, #1\right>}}
\newcommand{\braket}[2]{\ensuremath{\left< #1\, |\, #2\right>}}

\begin{document}
\vspace{0.01cm}
\begin{center}
{\Large\bf  Asymptotic dynamics, large gauge transformations and infrared symmetries } 

\end{center}

\vspace{0.1cm}

\begin{center}

{\bf Cesar Gomez}$^{a,e}$\footnote{cesar.gomez@uam.es} , {\bf Mischa Panchenko}$^{a}$\footnote{m.panchenko@campus.lmu.de}

\vspace{.6truecm}


{\em $^a$Arnold Sommerfeld Center for Theoretical Physics\\
Department f\"ur Physik, Ludwig-Maximilians-Universit\"at M\"unchen\\
Theresienstr.~37, 80333 M\"unchen, Germany}


%


{\em $^e$
Instituto de F\'{\i}sica Te\'orica UAM-CSIC, C-XVI \\
Universidad Aut\'onoma de Madrid,
Cantoblanco, 28049 Madrid, Spain}\\

\end{center}

\begin{abstract}
\noindent
Infrared divergences in QED and other theories with massless particles show that in such theories the $S$ matrix cannot be defined in the usual way. Typically, this is not viewed as a big problem since one is interested in cross sections, in which the divergences cancel. Recently, one particular type of divergences known as soft theorems was connected to a symmetry principle - the antipodal matching of large gauge transformations. However, there is a way to define an IR finite $S$ matrix in QED and similar theories by dropping the assumption of trivial asymptotic dynamics. In the present paper we investigate the role of soft theorems and invariance under large gauge transformations in the context of the finite $S$ matrix. Before doing so, the construction of asymptotic dynamics is reviewed and extended. The key results are that subleading soft factors can be included in a natural way in the asymptotic dynamics. Once this is done, soft modes decouple from the IR finite $S$ matrix and this decoupling, which can be understood as a spontaneously broken symmetry, is equivalent to the invariance under large gauge transformations (or in other words to the antipodal matching). To show this equivalence a special property for field operators at null infinity, i.e. for the program of asymptotic quantization, is assumed. Finally, we speculate about the modification of the decoupling of soft modes in the presence of black holes.
\end{abstract}

\newpage

\section{Introduction}

Typical $S$-matrix computations involving quantum field theories with massless particles lead to infrared divergent probability amplitudes. These divergences cancel in inclusive cross sections and are therefore often dubbed unphysical. Indeed, the amplitudes cannot be measured directly and experimentally relevant quantities are always IR finite. This has been known for a long time and consequently rather little attention was paid to the fact that even in such a simple theory as QED there is no $S$ matrix. \\

The authors of the present paper found the need to revise this old issue for two closely connected reasons. First: a particle with an energy $\omega$ typically interacts on time scales of $1/\omega$. Consequently, one would expect particles of zero energy to interact on infinite time scales only, or in other words not to interact at all. Thus, the probability of emitting or absorbing a photon of frequency $\omega$ in some physical process - say a scattering process - should vanish as $\omega\rightarrow 0$; infinitely soft photons should decouple from all processes. Instead however, using standard perturbation theory one finds that the probability for soft photons to participate in scattering diverges.
The second reason is a recent line of papers in which this divergent probability of soft emission/absorption was connected to invariance of the theory under large gauge transformations (LGT in the following) (insert citations). While it is always good to find new symmetries, one should be worried if these turn out to imply divergences: is there a symmetry reason for the non-existence of the $S$ matrix in QED and gravity?\\

The origin of IR divergent probability amplitudes is very clear: the usual definition of the $S$-matrix, or equivalently the LSZ formalism, assumes that the asymptotic dynamics of the theory is free. This assumption is clearly wrong when massless particles are around, hence the divergences. Several authors have tried to overcome this issue by finding the correct, non-free, asymptotic dynamics and defining a new, IR safe $S$ matrix \cite{FK, Zwanziger, Rohrlich}. In \cite{FK} it is mentioned that soft photons indeed decouple from the IR safe $S$-matrix of QED (no calculation of this result is presented and we are not aware of another source presenting it). In fact, this is still not quite true: as we show, they only decouple to leading order, resulting in non-divergent but still non-zero amplitudes. We show how to rectify this by including the subleading soft theorem into the asymptotic dynamics. This simultaneously solves an ambiguity in the construction of the latter. \\

With the IR safe $S$-matrix in our hands, the decoupling of soft photons of arbitrary direction can be interpreted as an infinite family of dynamical symmetries. Hence, it is natural to ask about the meaning and origin of these symmetries. It turns out, maybe a bit surprisingly, that the invariance of the theory under LGT follows from the decoupling of soft modes once the special role of null infinity in the framework of non-trivial asymptotic dynamics is recognized.\\

The main purposes of the present paper are the following: We improve on the methodology of \cite{FK} through a reexamination of the meaning of asymptotic dynamics and inclusion of subleading terms. Then a very simple calculation showing that soft photons decouple from the IR finite $S$-matrix is presented. After that, we show the connection between this decoupling and invariance of the theory under LGT. Before doing so, in order to avoid possible confusion, we stress the difference between invariance, matching and symmetry.\\

In spite of its intrinsic beauty, the interest in these results for QED or for linear gravity is mostly academic and invariance under LGT simply reflects the decoupling of soft modes. However, the whole story is richer and extends to other cases. One of the lessons lies in the way the theory implements its invariance under LGT through soft decoupling. The realisation of this symmetry is in the form of a spontaneously broken symmetry with a large vacuum degeneracy and with different vacua differing by a finite number of soft modes that play the role of Goldstone modes.  This is, as stressed in \cite{us1}, a generic phenomenon in theories with soft modes \footnote{In particular, in the simplest case of a free massless scalar field there is an infinite degeneracy and we can identify the corresponding soft modes playing the role of Goldstone bosons. The nontrivial aspect appears when the vacuum degeneracy coexists with nontrivial asymptotic dynamics. 
Similarly, in linearized gravity 
one obtains the formal vacuum degeneracy of Minkowski, as discussed in reference \cite{us1}.}.  Although this vacuum degeneracy corresponds to infinite entropy 
of asymptotically-Minkowski  vacuum, 
  the differences among the vacua cannot be resolved in {\it finite time}. This is a different way of  expressing the essence of the soft decoupling.\\
  
Several of the questions that we address here have recently been raised in \cite{amit} and  \cite{porrati}. While we largely agree with the conclusions reached there, the methodology that we use is quite different. We put a special emphasis to deriving all results from a priori arguments and carefully differentiating between the many different notions that necessarily appear in the context of non-trivial asymptotic dynamics. In particular, the special role of null infinity is recognized. 

A spacial slice approach to LGT was followed in \cite{Balachandran}, it would be interesting to see the precise connection to it. In \cite{Conde, Campiglia} the subleading soft theorem was put on a similar footing as the leading one, the same happens in the present paper. \\
  
The outline of the paper is as follows. First, we address the origin of infrared divergences and their disappearance once the appropriate asymptotic dynamics is chosen. Furthermore, we discuss ambiguities in the definition of the IR safe $S$ matrix and conjecture the correct choice of $S$. This will be done in sections \ref{AsHam}-\ref{section4}. The main result is that in the IR safe scenario zero energy modes decouple from the $S$-matrix. In section \ref{SymAndMatch} we discuss the connection between matching, invariances and symmetries. 
After that, we consider the role of LGT in the IR safe theory in section \ref{ScriExpansion}. We show that the charge of a LGT is simply an asymptotic soft photon. A key ingredient in this analysis dwells in the distinction between true asymptotic operators and what we will call \textit{free asymptotic operators}.  Finally, in section \ref{WhyScri} we discuss the issue of normalizability of scattering states. In the appendix \ref{app2} we will shortly review the standard treatment of infrared divergences and compare it to the present approach.
For concreteness, throughout the paper we focus on the case of QED. However, the arguments are in principle more general and can be applied to  different quantum field theories.

\section{The S matrix and asymptotic dynamics}\label{AsHam}
Generically, in theories with massless modes the asymptotic dynamics differs from the one defined by the free Hamiltonian. The correct identification of the asymptotic dynamics is necessary in order to define the $S$ matrix. In this section we shall briefly review the discussion on how to define the asymptotic dynamics. 
\subsection{What is an S matrix?}\label{Smat}
In principle, a scattering amplitude is just the overlap of two states. Typically, one is interested in the overlap of scattering states, i.e. true eigenstates of the full Hamiltonian. Since these are usually unavailable, one resorts to \textit{descriptor states} and an operator on such states - the $S$ matrix - that describes scattering and can be expanded in a perturbative series.  The descriptor states and the $S$ matrix are ideally obtained through the following procedure:
\\

Let $W(t)=\ee^{-\iu H t}$ be the full time evolution operator and let $U(t)$ be some other unitary evolution. We can match the dynamics governed by the two different time evolutions at any fixed point of time. For that let $\ket{g}, \ket{f_g}$ be two different states such that
\begin{equation}
W(t_1)\ket{g}=U(t_1)\ket{f_g} \Longleftrightarrow \ket{g}= \Omega(t_1)\ket{f_g} 
\end{equation}
where  $\quad \Omega(t):=W(-t)U(t)$. Obviously this can be done for any $U(t)$. We could now try to define an $S$ matrix via
\begin{equation}
\braket{h_{out}}{g_{in}}=\lim_{t_\pm\rightarrow\pm\infty}\bra{f_h}\Omega^\dagger(t_+)\Omega(t_-)\ket{f_g} = \bra{f_h}S\ket{f_g}
\end{equation}
with $S=\Omega^\dagger(\infty)\Omega(-\infty)=\Omega_+^\dagger\Omega_-$. For this procedure to work, the convergence of the M\o ller operators $\Omega(t)$ is of crucial importance. Indeed, it is only this need of convergence what forbids us to use an arbitrary $U(t)$ in the definition of $S$. We shall call the ``correct" evolution operator $U_{as}(t)$. \\

The convergence of $\Omega_\pm(t)$ is not the only criterion for the correct definition of $U_{as}$. Since we want to make contact with real experiments, $U_{as}$ should be such that for chosen descriptor states $\ket{f_g}$ the asymptotics, as governed by $U_{as}$, corresponds to the asymptotics of the ``real" state $\ket{g_{in/out}}$ as governed by $W$, i.e. 
\begin{equation}\label{mapping1}
\lim_{t_\mp\rightarrow\infty}\left( W(t)\ket{g_{in/out}}-U_{as}(t)\ket{f_g} \right)=0 \quad \Longleftrightarrow \quad \Omega_\pm\ket{f_g}=\ket{g_{in/out}}
\end{equation}
where the convergence is with respect to an appropriate norm. If that is the case, one can be sure that $\Omega(t)$ has strong limits and that the scattering amplitude $\bra{f_h}S\ket{f_g}$ is equal to the physically meaningful quantity $\braket{h_{out}}{g_{in}}$. For instance, in quantum mechanical potential scattering, descriptor states are eigenstates of the momentum operator $\ket{p}$ and a correctly chosen $U_{as}(t)$ must lead to 
\begin{equation}\label{mapping2}
\lim_{t_\mp\rightarrow\infty}\Omega(t)\ket{p}=\ket{p_{in/out}},
\end{equation} 
where $\ket{p_{in/out}}$ are the (typically known)  in/out stationary scattering states. Only in that case does the $S$ matrix reproduce the usual formula for the scattering amplitude.  Note that this $S$ matrix can be written as $W_{I_{as}}(\infty,-\infty)$ where the interaction picture is defined with the $U_{as}$, i.e. 
\begin{align}
\ket{\psi(t)}_{I_{as}} &=U_{as}^{-1}(t)\ket{\psi(t)}_S \Longrightarrow \iu \partial_t\ket{\psi(t)}_{I_{as}}\xrightarrow{\,\abs{t}\rightarrow\infty\,} 0 \\
A_{I_{as}}(t)&=U_{as}^{-1}(t)\, A\, U_{as}(t)\, .
\end{align}

For short range interactions (i.e. only massive particles) the asymptotic dynamics is governed by the (renormalized) free Hamiltonian, hence $U_{as}(t)=\ee^{-\iu H_0 t}$ is a good choice and the correspondingly defined $S$ matrix leads to correct results. For long range interactions (like QED) this is not the case and one must choose a different $U_{as}(t)$ as was first shown by Dollard \cite{Dollard} for Coulomb potential scattering and subsequently applied to QED by Kulish and Faddeev in \cite{FK}. In what follows, we shall show how to find the asymptotic dynamics and how to construct the correct $S$ matrix from it.

\subsection{Finding the asymptotic dynamics}
Finding the correct asymptotic dynamics for theories with long range interactions is a difficult task. While the descriptor states can still be conveniently chosen to be eigenstates of the free Hamiltonian, one usually does not have any kind of expression for the true scattering states (or equivalently incoming/outgoing particles). In principle, if the scattering states $\ket{p_{in/out}}$ are known, the goal is to find an operator $U_{as}(t)$ that fulfills two requirements:
\begin{enumerate}
\item The M\o ller operators $\Omega_{\pm}=\lim\limits_{t\rightarrow\pm\infty} \ee^{-\iu \,H\,t} U_{as}(t)$ exist, or equivalently the $S$-matrix $\Omega_+^\dagger\Omega_-$ is finite on the Hilbert space of the descriptor states.
\item The operators $\Omega_{\pm}$ map descriptor states to true scattering states, i.e. $\Omega_\pm \ket{p}=\ket{p_{out/in}}$.
\end{enumerate}
In most situations the scattering states are unknown and the second requirement is ignored, i.e. one is happy to obtain just some finite $S$ matrix. As we will see, it is the second requirement which forces us to include the subleading soft factors into the asymptotic dynamics. \\

In the literature there are (as far as we know) two ways for obtaining a $U_{as}(t)$ that fulfills the first requirement, both generically leading to the same results. As we will need both of them, we shall give a short summary here.

\subsection{Ansatz I: asymptotic dynamics from leading order terms}
This trick for finding $U_{as}(t)$ was (to our knowledge) first used in \cite{FK}. It goes as following: first use the following trivial identity
\begin{equation}
H=H_0+V = H_0+\ee^{-\iu H_0t}\,V_I(t) \,\ee^{\,\iu H_0t},
\end{equation}
where the subscript $I$ refers to the usual interaction picture. Now compute $V_I(t)$ to highest order in $t$ and neglect subleading terms; call the resulting operator $V_{I \, as}(t)$. Finally define $V_{as}(t):=\ee^{-\iu H_0t}\,V_{I \,as}(t) \,\ee^{\iu H_0t}$ and
\begin{equation}\label{Has}
H_{as}(t):=H_0+V_{as}(t).
\end{equation}
By construction, $H_{as}(t)$ will converge to $H$ for large $t$. Hence, the asymptotic dynamics in the Schr\"odinger picture can be described as:
\begin{equation}
H_{as}(t)\ket{\psi}_S\xrightarrow{\,\abs{t}\rightarrow\infty\,}\iu \partial_t \ket{\psi}_S
\end{equation}
or in other words, for asymptotic times:
\begin{equation}\label{Uas}
\iu \partial_t U_{as}(t)=H_{as}(t) U_{as}(t).
\end{equation}
This leads to a differential equation for $U_{as}(t)$. The resulting Dyson series for $U_{as}(t)$ can now be computed order by order. Nevertheless, there is a subtlety: equation \ref{Uas} is only meaningful for asymptotic times and hence we cannot demand $U_{as}(t_0)=\mathbb{1}$ for some $t_0$. In other words, the general solution for $U_{as}(t)$ takes the form
\begin{equation}\label{Uassol}
U_{as}(t)=\mathcal{T}\,\ee^{-\iu \int_{t_0}^t H_{as}(\tilde{t}) d\tilde{t}} \cdot C,
\end{equation}
where $C$ is some time-independent unitary operator.\\

The unknown $C$ has to be fixed by the choice of descriptor states and the condition (\ref{mapping1}), we will come back to its role in section \ref{C}.

\subsection{Ansatz II: asymptotic dynamics from classical solutions}
This method for determining $U_{as}(t)$ was used e.g. by \cite{Zwanziger} for QED. Here, instead of finding $U_{as}$ directly, one makes an Ansatz for some operators in the Heisenberg picture to fulfill a ``classically inspired" time evolution for large times. Then, by plugging this Ansatz into the e.o.m. for other operators in the Heiseberg picture, the asymptotic dynamics of the entire system is solved. Alternatively, one can look for an evolution operator $U_{as}(t)$ that gives the classically inspired dynamics. In this procedure, the above unknown operator $C$ is fixed by subleading terms in the Ansatz. This might sound a bit obscure at first, but is in fact both conceptually and technically quite simple. In order for the reader to become familiar with the two ways of finding asymptotic dynamics, we will treat a completely understood example below - the Coulomb potential scattering in quantum mechanics.

\section{Worked out example: Coulomb scattering}
We consider the Coulomb potential in three dimensions:
\begin{equation}
H=\frac{p^2}{2m}+\frac{g}{r} = H_0+V.
\end{equation}
\subsection{Ansatz I}
Let us follow Ansatz I first. With 
\begin{equation}
\vec{r}_I(t) = \vec{r}+\vec{p}\frac{t}{m}\approx \vec{p}\frac{t}{m} \quad \text{ for large }\abs{t}.
\end{equation}
we find:
\begin{equation}
V_{as}(t)= \frac{g\,m}{\abs{t} p}
\end{equation}
Then equations \ref{Has} and \ref{Uassol} result in 
\begin{equation}\label{UCoul}
U_{as}(t)=\ee^{-\iu\left(\frac{p^2}{2m}(t-t_0)+\frac{g\,m}{p}\text{sign}(t)\log(\,\abs{\frac{t}{t_0}}) \right)} \, C.
\end{equation}

\subsection{Ansatz II}\label{ansatz2}
Now to the Ansatz inspired by classical Coulomb scattering. We want to find approximate solutions to the classical e.o.m.
\begin{equation}
\dot{\vec{r}}=\frac{1}{m}\vec{p} \quad, \quad \ddot{\vec{r}}=\frac{g}{m\,r^3}\vec{r}
\end{equation}
for large $\abs{t}$ and $r$. The momentum $\vec{p}$ must be conserved to leading order in $t$ in this regime. One can check that the following expressions fulfill these requirements
\begin{align}
&\vec{r}(t)\approx \vec{p}\frac{t}{m}-\vec{p}\frac{g\,m}{p^3}\text{sign}(t)\log(\abs{t}) +\text{const.} \\
&\vec{p}(t) \approx \vec{p}
\end{align}
One can include a constant term since the solution is valid only for very large $r$ and $t$ anyway. Note that the equation for $r$ is simply the movement of free particles corrected by a subleading (in $\abs{t}$) term coming from the Coulomb interaction. The idea is now to assume that in the Heisenberg picture the operators $\vec{r}(t)$ fulfills the same equation for asymptotic times (we do not write hats over operators, it should be clear from the context when the variables are classical and when quantum). Then the asymptotic time evolution can be found from
\begin{align}
&U_{as}(t)^\dagger\, \vec{p}\, U_{as}(t) = \vec{p} \\
&U_{as}(t)^\dagger \, \vec{r} \, U_{as}(t) = \vec{p}\frac{t}{m}-\vec{p}\frac{g\,m}{p^3}\text{sign}(t)\log(\abs{t}) +\text{const.} 
\end{align}
It is easy to see that \ref{UCoul} gives precisely that (where $C$ must commute with $p$), hence both Ansatzes lead to the same result.

\subsection{The role of C}\label{C}
The operator $C$ is mostly ignored in discussions of asymptotic dynamics. In the present paper, however, it plays a central role and this section is therefore quite important. $C$ can be found by fixing the descriptor states for Ansatz I or by fixing the constant term in Ansatz II. Let us see how the first part works. For the Coulomb potential the scattering states $\ket{\vec{p}_{in/out}}$ are well known, see e.g. \cite{Shankar}. As descriptor states we choose, as usual, the eigenstates of $\vec{p}$, call them $\ket{\vec{p}}$. With the help of $U_{as}(t)$ we define the (now converging) M\o ller operators as in section \ref{Smat}. $C$ is now fixed by the requirement 
\begin{equation}\label{requirement}
\Omega_{\pm}\ket{\vec{p}}=\ket{\vec{p}_{out/in}}
\end{equation}
If we ignore that requirement and e.g. set $C = 1$, the M\o ller operators defined through $U_{as}(t)$ will still converge and the S-matrix will be finite. However, the matrix element $\bra{p'} S \ket{p}\neq \braket{p'_{out}}{p_{in}}$ and calculations using this $S$ matrix will in general lead to wrong results. The choice of $C$ can be viewed as a renormalization of the $S$ matrix, i.e. a modification of the LSZ prescription for asymptotic states. Since the asymptotic dynamics cannot depend on $t_0$, $C$ also eliminates the dependence on $t_0$ in $U_{as}(t)$. See \cite{Rohrlich} for a nice discussion of why this must happen. It seems that the unphysical and non-existent dependence on $t_0$ plays a crucial role in the work \cite{amit}, we strongly disagree with their point of view on that.\\

It turns out that the with the right choice for $C$\footnote{In fact, one has to choose two different $C$'s - one for future times and one for fast times. They merely differ by a sign, in order to shorten the presentation we ignore this subtlety.} the asymptotic evolution is given by:
\begin{equation}
U_{as}(t)=\ee^{-\iu\left(\frac{p^2}{2m}t+\frac{g\,m}{p}\text{sign}(t)\log(\,\abs{t} \frac{p^2}{2m}) \right)}.
\end{equation}
This was proven in \cite{Dollard}, where it was shown that with this choice equation \ref{requirement} holds. Note that even in this simple case the operator $C$ is nontrivial!\\

Now as for the second Ansatz, we see that the correct asymptotic dynamics leads to
\begin{equation}
\vec{r}_{as}(t)=\vec{p}\frac{t}{m}-\vec{p}\frac{g\,m}{p^3}\text{sign}(t)\log(\abs{t}\frac{p^2}{2m})+\vec{r}+2\vec{p}\frac{g\,m}{p^3}.
\end{equation}
The asymptotic expression for $r(t)$ has a constant piece containing the operator $\vec{p}$. While we have no intuition for the meaning of this piece in potential scattering, its analogue in QED will be played by universal subleading soft factors.

\section{The infrared safe S-matrix in QED}\label{section4}
In this section the program of finding the correct asymptotic dynamics is applied to QED. We largely follow the ideas of \cite{FK} and \cite{Zwanziger}, although with a different emphasis and some extension. The main point of the present discussion is the decoupling of soft modes from the IR safe $S$-matrix. For the modes to decouple completely, subleading effects have to be taken into account. This is where the previously ignored operator $C$ enters the stage. Admittedly, the presented construction is a bit ad hoc - although a classical argument for the inclusion of subleading terms is given, the main reason for them is the demanded decoupling. It would be nice to find a more natural argument that leads to the subleading dressing. The section is structured as follows: in \ref{softphot} the leading and subleading soft photon theorems are reformulated as operator identities. Then, in \ref{asdym}, after a short recap of known results the decoupling of soft photons is demonstrated to leading order. Our approach here differs from the usual. We do not modify the descriptor states, as was done in previous work, but rather treat the dressing operators with more care. Certainly, the free Fock space has to be extended in some way in order to make sense of the dressing, but the space of descriptor states should still be unchanged if we are to follow the philosophy of Dollard in \cite{Dollard}. In particular, the IR safe $S$ matrix should be a well defined and finite operator on the free Fock space. In the remaining sections \ref{subleading} and \ref{explanation} the subleading effects are taken into account and a short note about soft theorems is given.
\subsection{The soft photon theorem in QED}\label{softphot}
Assuming trivial asymptotic dynamics in QED and performing computations with the usual Dyson $S$ matrix, let us call it $S_D$, leads to the well known problem of IR divergences. One very particular type of such divergences is known under the name \textit{soft photon theorem}. It is usually stated as an equation for matrix elements of $S$, namely

\begin{align}
\lim_{k^0\rightarrow 0}\bra{p'_i}a_\mu(k)S_D\ket{p_j}= &\left( \lim_{k^0\rightarrow0}\sum \frac{e_i}{p'_i\cdot k}\left(\,p'_\mu-\iu \,k^\nu \,J_{\mu\nu}(p')\right)- \text{ same for p } \right) \cdot  \nonumber \\
&\bra{p'_i}S_D\ket{p_j}
\end{align}
where we have included the non-divergent subleading part with the angular momentum operator
$$J_{\mu\nu}(p)= \iu \left( p_\mu \frac{\partial}{\partial p^\nu}-p_\nu \frac{\partial}{\partial p^\nu} \right).$$ 

The operators $a_\mu$ create the Fock space of all polarizations, the physical Fock space being a subspace of it.\\

Let us only consider matrix elements with no arbitrarily soft photons in $\ket{p_j}$ and $\ket{p'_i}$ . Then we get 
$$
\lim_{k^0\rightarrow 0}\bra{p'_i}a^\mu(k)S_D\ket{p_j}=\lim_{k^0\rightarrow 0}\bra{p'_i}[a^\mu(k),S_D]\ket{p_j}.
$$
Since this is true for all matrix elements, we can promote the soft photon theorem into an operator equality:
\begin{equation}\label{soft}
\lim_{k_0\rightarrow 0}[a_\mu(k),S_D] = \lim_{k^0\rightarrow 0} e \int \frac{1}{p\cdot k}\,[\left( \,p_\mu -\iu \,k^\nu \,J_{\mu\nu}(p) \, \right) \,\rho(p) \, , \, S_D] \,d^3p.
\end{equation}
where $\rho(p)$ is the charge density operator. Obviously, the above equation can be written as a symmetry,  i.e. as
\begin{equation}
 [Q,S_D]=0
\end{equation}
where the conserved charge is
\begin{align}\label{soft2}
 Q=&\lim_{k_0\rightarrow 0} \,a_\mu(k) - e \lim_{k_0\rightarrow 0}\int \frac{d^3p}{p\cdot k}\,\left(p_\mu \,-\iu \,k^\nu \,J_{\mu\nu}(p)\right)\rho(p) \nonumber \\
 &=: Q_{soft}+Q_{hard}.
\end{align}

Note that this is not quite a symmetry yet since the charge is not hermitian. If one wants to make the statement about hermitian charges only, one can use the operators $a+a^\dagger$ and $\iu(a-a^\dagger)$. They will pick out only the leading and subleading parts of the soft theorems respectively.

\subsection{Asymptotic dynamics and soft decoupling in QED to leading order}\label{asdym}
The reason for the IR divergences in the Dyson S matrix is the wrong assumption of trivial asymptotic dynamics. In fact, for QED a completely analogous analysis to the one presented for the Coulomb potential can be done. As was shown in \cite{FK} and \cite{Zwanziger} (although $C$ was set to the identity there), the real asymptotic dynamics is governed by:

\begin{equation}
U_{as}(t)=\ee^{-\iu H_0 t} \ee^{R(t)} \ee^{\iu \Phi(t)} \, C
\end{equation}
with 
\begin{align}
&R(t)=\frac{e}{(2\pi)^3}\int \frac{p^\mu}{p\cdot k}\,\rho(p)(a^\dagger_\mu\ee^{\iu\frac{k\cdot p t}{p_0}}-h.c.)\frac{d^3k}{2 k_0} d^3p \\
& \Phi(t) \sim \int \, : \rho(p)\, \rho(q) : \frac{p\cdot q}{((p\cdot q)^2-m^4)^\frac{1}{2}} \,\text{sign}(t) \ln (\,\abs{t}) \,d^3p\,d^3q
\end{align}

There, for the Ansatz of the type \ref{ansatz2} one assumes a classical form for the asymptotic current in the Heisenberg picture:
\begin{equation}\label{jas}
J_\mu^{as}(x)=\int d^4p\, \rho(p) \int\limits_{-\infty}^\infty d\tau\,\delta^4(x-p\tau),
\end{equation}
this is the current of particles of momentum p flying on straight lines through the origin. This leads to
\begin{equation}\label{aas}
A_\mu^{as}(x)=A_\mu^{in}(x)+\int d^4y\, \Delta^{ret}(x-y) J_\mu^{as}(y),
\end{equation}
which is the incoming electromagnetic field and the Lienard-Wiechert field of the asymptotic charged particles; and then ultimately to the same $U_{as}(t)$.\\

Note that the time-dependent phase $\ee^{\iu\frac{k\cdot p t}{p_0}}$ in the dressing operator $R(t)$ suppresses the contribution of any finite $k$ modes in the $\abs{t}\rightarrow \infty$ limit. Hence, the dressing is purely soft. The phase operator $\Phi(t)$  is the direct analogue of the phase for non-relativistic Coulomb scattering, it commutes with photons and is irrelevant for the present discussion. We will ignore it in the following.\\

Since the analogue of the Coulomb potential scattering states is unavailable in QED, we need, in order to fix $C$, to come up with a physical argument. Before doing that and in order to highlight the physical meaning of $C$ let us just take for the moment $C=1$.\\

The definition of $U_{as}$ implies the following $S$ matrix as acting on the free Fock space (where we ignore the phase operator):
\begin{equation}
S_{KF}=\lim_{t_\pm\rightarrow\pm\infty} \, \ee^{R(t_+)^\dagger}\, S_D(t_+,t_-) \,\ee^{R(t_-)} =: \lim_{t_\pm\rightarrow\pm\infty} S_{KF}(t_+,t_-)
\end{equation} 
where $KF$ stands for Kulish-Faddeev and $S_D$ is the standard Dyson $S$ matrix in the free interaction picture, i.e. $S_D(t_+,t_-)=U_I(t_+,t_-)$.\\

The leading order of the soft photon theorem, as stated in (\ref{soft}), implies that 
\begin{equation}\label{decoupling1}
\lim_{k\rightarrow 0} [k_0\,a^\mu(k),S_{KF}]\, = 0 .
\end{equation}
This relation can be easily proved. First note that
\begin{small}
\begin{align*}
&[a^\mu(k),S_{KF}(t_+,t_-)]=\\[2mm]
 [a^\mu(k),\ee^{R(t_+)^\dagger}]\, S_D(t_+,t_-) \,\ee^{R(t_-)} +  & \ee^{R(t_+)^\dagger}\,[a^\mu(k), S_D(t_+,t_-)] \,\ee^{R(t_-)} + \ee^{R(t_+)^\dagger}\, S_D(t_+,t_-) \,[a^\mu(k),\ee^{R(t_-)}] 
\end{align*}
\end{small}
We can now use standard formulas for displacement operators and take the $k\rightarrow0$ limit before the $t_\pm\rightarrow\pm\infty$ limit finding (schematically):
\begin{align}
&\quad\lim_{k_0\rightarrow 0} [k_0\,a^\mu(k),S_{KF}]  = \nonumber\\
 \ee^{R^\dagger(\infty)} \lim_{k^0\rightarrow 0} \, k_0\, &\left( e \int \frac{p^\mu}{p\cdot k}\,[-\rho(p),S_D] dp+[a_\mu(k),S_D] \right) \ee^{R(-\infty)}.
\end{align}
The soft photon theorem \ref{soft} implies that the expression in between the dressing operators vanishes and hence equation (\ref{decoupling1}) follows. It is important to notice that the limits of $k_0\rightarrow 0$ and $t_\pm\rightarrow \pm\infty$ do not commute, therefore the above calculation should be rather viewed as a guide to a more careful and rigorous one. However, the conclusion of the more careful analysis (where $t_\pm$ is kept large but finite with the limit being taken at the very end) is the same. In general, for all the following calculations of this type we will take $k_0\rightarrow 0$ before the infinite time limit.\\

Equation (\ref{decoupling1}) implies that amplitudes with participation of soft photons are no longer IR divergent. The next question is how to fix the constant operator $C$ and what is its meaning at the level of soft theorems. The answer is that $C$ contains information about the {\it subleading part of the soft theorems}. 

\subsection{Including the subleading dressing}\label{subleading}

Once we include $C$ the $S$ matrix becomes
\begin{equation}
S_{KF} =\lim_{t_\pm\rightarrow\pm\infty}C^\dagger \, \ee^{R(t_+)^\dagger}\, S_D(t_+,t_-) \,\ee^{R(t_-)} C,
\end{equation} 
In order to identify $C$ we shall use two different arguments
 which lead to the same conclusion:
\begin{enumerate}
 \item The insertion of soft photons must be a symmetry of scattering, i.e. soft photons decouple to all orders and not just to the leading one. \footnote{A similar philosophy was followed in reference \cite{porrati}}
 \item The asymptotic expression for the electromagnetic field $A_\mu(\vec{x})$ states must reduce to its ``classical value''.
\end{enumerate}

Let us define
\begin{equation}
\ee^{R(t)}C=:\ee^{\tilde{R}(t)}
\end{equation}
and make the following Ansatz:
\begin{equation}
\tilde{R}(t)=R(t)+\left( \int\, d^3p\, d^3k\, a_\mu(k) C^\mu(k,p) -h.c. \right)
\end{equation} 
where we impose that $C^\mu$ are Lorentz covariant operators that do not contain photon operators, as well as the condition
\begin{equation}
[C^\mu(p,k),\rho(p')]=0 \qquad \forall k,p\neq p'.
\end{equation} 
In other words, we demand that the asymptotic states $U_{as}(t)\ket{p}$ are ``coherent" (in the sense of section \ref{ScriExpansion} below) and do not involve other momenta $p'\neq p$ even if $C$ is chosen to be nontrivial.\\

Then we obtain, in analogy to the leading order calculation:
\begin{align}
&\lim_{k_0\rightarrow 0} [a^\mu(k),S_K]  = \nonumber \\
&\ee^{\tilde{R}(\infty)^\dagger} \lim_{k_0\rightarrow 0}\left( e \int \frac{p^\mu}{p\cdot k}\,[-\rho(p)+ C^\mu(k,p),S_D] dp+[a_\mu(k),S_D] \right) \ee^{\tilde{R}(-\infty)},
\end{align} 
Again the limit $k_0\rightarrow 0$ must be taken before the limit $t_\pm\rightarrow \pm\infty$. Note that if we were to take the limit at the beginning of the calculation, the dressing operator $\ee^{R(t)}$ would appear to become trivial (as was already noted in \cite{FK}).\\ 

Now let us fix $C$ by demanding that soft photon insertions are a symmetry of the infrared finite $S$ matrix, i.e. 

\begin{equation}\label{fulldecoupling}
\lim\limits_{k^0\rightarrow 0} [a^\mu(k),S_{KF}]=0.
\end{equation}
Note that we need this to hold if normalized zero energy photons are to have zero probability of being created or absorbed in physical processes, see the discussion in the introduction. Decoupling to leading order is not enough for this. This assumption together with (\ref{soft}) implies that (to order $\mathcal{O}(k)$):
\begin{align}\label{R(t)}
\tilde{R}(t)=R(t)+\frac{\iu}{(2\pi)^3}\int\, d^3p\, d^3k\,\frac{1}{2k_0} \,\frac{h(k)}{p\cdot k}\left(a_\mu(k)+a^\dagger_\mu(k)\right) k_\nu \, J^{\mu\nu}(p)\rho(p) = \nonumber \\
\frac{e}{(2\pi)^3}          \int        \frac{1}{p\cdot k}    \left(\ee^{\iu\frac{k\cdot p t}{p_0}}\,p^\mu\,-\iu \, h(k)\, k_\nu \, J^{\mu\nu}(p) \right)\rho(p)     \,  a^\dagger_\mu(k) \,     \,    \frac{d^3k}{2 k_0} d^3p -h.c.
\end{align}
Here $h(k)$ is some window function with $h(0)=1$ which we cannot determine a priori. Realistic experiments with imperfect detectors can determine $h$ up to a dependence on the detector resolution. In any case, we will only perform calculations in the $k\rightarrow 0$ limit, hence the exact form of $h(k)$ will never enter.\\
In summary, we observe that the subleading soft photon theorem fixes the factor $C$ of the dressing operator. In the following we will omit the tilde in the notation and whenever we say $R(t)$ the full operator $\tilde{R}(t)$ is meant.\\

Let us now discuss how $C$ can be constructed from classical arguments. We have already seen in equation \ref{aas} that the leading order asymptotic dynamics gives for the potential $A_\mu^{as}$ the electromagnetic field of point particles flying on straight lines through the origin. The subleading dressing gives the following addition to it:
\begin{align}
\tilde{A}_\mu^{as}(x)=A_\mu^{as}(x)+A_\mu^{sub}(x) \nonumber \\
A_\mu^{sub}(x) = \frac{\iu}{(2\pi)^3} \int \frac{d^3k}{2k_0} \,d^3p\, h(k) \, k^\nu J_{\mu\nu}(p) \rho(p) \frac{\ee^{\iu \,px}}{p\cdot k} +\, h.c.
\end{align}
The asymptotic electromagnetic field operator (in the Feynman gauge) consists of the incoming field, the Lienard-Wiechert field  and a part depending on the angular momentum of the asymptotic charged particles. In other words, it corrects the previous expression for the presence of non-zero angular momentum which should certainly influence the electromagnetic field. Note that the asymptotic current is unchanged since 
\begin{equation}
\Box\, A_\mu^{sub}=0.
\end{equation}

\subsection{Soft theorems and dressing}\label{explanation}
In order to avoid a possible misunderstanding, we think it is worth to make explicit the following straightforward result. Whenever one has \textit{any kind of soft theorem} in the IR divergent theory, i.e. some operator $Q_{soft}$ consisting out of purely soft photons and a corresponding hard part that is consistent with the soft photon theorem so that $Q=Q_{soft}+Q_{hard}$ commutes with $S_D$ - the following equation holds:
\begin{equation}\label{soft3}
 \ee^{R(t)}\,Q_{soft}\,\ee^{R(t)\dagger}= Q_{soft}+Q_{hard}.
\end{equation} 
This follows from the usual properties of the coherent dressing as a displacement operators and from formulas (\ref{R(t)}) and (\ref{soft2}).

\section{Symmetries and matching conditions}\label{SymAndMatch}
\subsection{Two different S matrices}
In order to clarify the following discussion, let us introduce some notation and distinctions. For scattering the relevant quantity is the amplitude $\braket{p_{out}}{p'_{in}}$ where $\ket{p_{in}},\ket{p_{out}}$ corresponds to stationary scattering states, i.e. true eigenstates of the full Hamiltonian. There are two different strategies for describing a scattering process:
\begin{enumerate}
 \item To use ``free" descriptor states $\ket{p}$ and M\o ller operators $\Omega_{\pm}$ that map them into true scattering states, i.e. $\Omega_\pm\ket{p}=\ket{p_{out/in}}$. In this case the relevant Fock space is the free Fock space, $\F_{free}$, and the amplitude becomes:
 \begin{equation}
 \braket{p_{out}}{p'_{in}}=\bra{p}S\ket{p'} \quad,\quad S = \Omega_+^\dagger \Omega_- \,.
 \end{equation}

\item Use either $\ket{p_{in}}$ or $\ket{p_{out}}$ to construct the Fock spaces $\F_{in/out}$. For concreteness, let us work with $\F_{in}$. In that case the states $\ket{p_{out}}$ must be described as a superposition of in states $\ket{p_{in}}$. We introduce a \textit{different} $S$ matrix, say $\tilde{S}$, via $\bra{p_{out}}=\bra{p_{in}}\tilde{S}$. The scattering amplitude then becomes:
 \begin{equation}\label{Stilde}
 \braket{p_{out}}{p'_{in}}=\bra{p_{in}}\tilde{S}\ket{p'_{in}} \quad,\quad \tilde{S} = \Omega_-\Omega_+^\dagger \,.
 \end{equation}
\end{enumerate}

If the assumption of trivial asymptotic dynamics is made, the M\o ller operators are defined via
\begin{equation}
\Omega_\pm=\lim_{t\rightarrow\pm\infty}\ee^{\iu \,H\,t}\,\ee^{-\iu\,H_0\,t} \Longrightarrow S = U_I(\infty,-\infty).
\end{equation}

\subsection{Symmetries and conservation laws}
It is easy to find useful expressions for  conserved charges with the first approach to scattering. Indeed, conserved charges $Q$ typically commute with the M\o ller operators and with $S$. Therefore, the action of the charge on $\ket{p_{in/out}}$ is fully determined by the action of $Q$ on $\ket{p}$:
\begin{equation}
Q\ket{p_{out/in}}=\Omega_\pm \,Q \,\ket{p}.
\end{equation}
Hence, for all calculations it is enough to know how $Q$ acts on descriptor states. Since the descriptor Fock space is built out of field operators at a fixed time $t=0$, explicit expressions for $Q\ket{p}$ can be found immediately - one just needs to express $Q$ in terms of operators at $t=0$.\\

For the second approach the situation is slightly more complex. In fact, the charge operators $Q$ are normally  not expressed in terms of the asymptotic fields.  In particular, let us consider generic operators $Q$ that \textit{do not} commute with $\tilde{S}$ and try to get useful information out of them. Later on we shall see how this procedure works for LGT.\\

First note that for any operator $Q$ acting on the ``free" Fock space of descriptor states we can define the corresponding in and out operator as
\begin{equation}\label{Qin}
Q_{in}:=\Omega_-\,Q\,\Omega_-^\dagger \quad, \quad Q_{out} = \Omega_+ \, Q \,\Omega_+^\dagger = \tilde{S}^\dagger\,Q_{in}\,\tilde{S}.
\end{equation}
If $Q$ is expressed in terms of creation/annihilation operators of free states, say $a(k)$, then $Q_{in/out}$ is the same operator with $a(k)$ replaced by $a_{in/out}(k)$ - the creation/annihilation operators of the true scattering states.

By definition we get
\begin{equation}\label{Qout}
\tilde{S}\,Q_{out}-Q_{in}\,\tilde{S}=0.
\end{equation}
Note that the above equation (\ref{Qout}) is a trivial identity holding for all $Q$. \\

Let us now consider how LGT become a symmetry. For that let us construct the Fock spaces of incoming and outgoing states separately $\F_{in}=\F_{out}$. For each construction we can get explicit expressions for the charges of LGT as acting on $\ket{p_{in}}\in \F_{in}$ and $\ket{p_{out}}\in \F_{out}$, let us call them $Q_-(\varepsilon_-)$ and $Q_+(\varepsilon_+)$ where $\varepsilon_\pm$ are arbitrary functions on a sphere. Note that the explicit action of $Q_-$ is known on in-states and that of $Q_+$ is known on out-states only. Let us fix some function $\varepsilon_-$. In principle there is no guarantee that there exists an $\varepsilon_+$ such that
\begin{equation}\label{LGTinvarience}
Q_-(\varepsilon_-)=Q_+(\varepsilon_+)
\end{equation}
is fulfilled. What is meant by \textit{invariance of scattering under LGT} is precisely the existence of a $\varepsilon_+$ given an $\varepsilon_-$ such that (\ref{LGTinvarience}) is fulfilled, i.e. the existence of a {\it matching}.  Once the matching is known, nontrivial results can be derived from it, e.g. by using equations (\ref{Qout}) and (\ref{Qin}). In references \cite{strominger1, strominger2, strominger3} antipodal identification of the gauge parameters $\varepsilon_+$ and $\varepsilon_-$ was shown to be equivalent to the soft photon theorem and therefore it proves the equivalence between the soft photon theorem and the invariance under LGT. It is, however, important to notice that what is meant here by invariance under LGT does not necessarily correspond to symmetries in the conventional sense. Indeed, nowhere in the calculation was demanded that $[Q,\tilde{S}]=0$ for any charge $Q$. 

\subsection{Large gauge transformations and fluxes}
As we have pointed out in the introduction, in those cases where the asymptotic dynamics is nontrivial, gauge transformations that are non vanishing at infinity lead to nontrivial transformations of asymptotic states. The way these transformations become a real symmetry of the theory is through the decoupling of soft modes underlaying the infrared finiteness of the theory. 

In classical physics we are used to associate real symmetries with conserved Noether charges. These charges are defined by integrating the corresponding conserved current on a space-like hypersurface and the charges are conserved in time. However, if instead of a space-like surface we use a null Cauchy surface what we define by integrating the corresponding conserved current is the flux through the null surface, we shall denote these fluxes {\it Noether fluxes}. 

In the case of the soft decoupling symmetry of the $S$ matrix the relevant {\it Noether flux} associated with the symmetry should count the flux of soft modes going through  null infinity for any given asymptotic state defined by the momentum and charges of  ingoing and outgoing charged particles. 

The soft decoupling symmetry of the infrared finite $S$ matrix implies the decoupling of this flux of soft modes. The way this is done at the level of the infrared finite $S$ matrix is by effectively reabsorbing  the flux of soft modes into the asymptotic coherent state dressing. Equivalently, this decoupling can be achieved if the $S$ matrix, without the dressing, commutes with the Noether flux of soft modes.

Although the {\it Noether flux} of soft modes is the relevant charge to implement the $S$ matrix symmetry associated with decoupling of soft modes, we can define standard charges on space-like hypersurfaces that correspond to adding a soft mode of certain polarisation and direction. These are actually the charges that we can think of as being spontaneously broken and creating an effective Goldstone boson. The quantum mechanical aetiology of these charges lies again in the infrared physics, namely in the dressing by a coherent state of {\it infinite} number of soft modes. In summary, soft decoupling is associated with LGT Noether fluxes while vacuum degeneracy is associated with standard charges on space-like hypersurfaces.

A typical example of LGT with important physical implications appears in those gauge theories where topologically nontrivial gauge transformations exist. In those cases it is the nontrivial topology that provides the possibility of LGT. In the case of infrared symmetries, what replaces the nontrivial topology are the infrared divergences. 

\subsection{The soft photon theorem from matching}
In reference \cite{strominger1} it was shown, assuming trivial asymptotic dynamics, that the charges for LGT in massless QED can be written in terms of out/in operators as
\begin{align}\label{LGTcharges}
Q_+(\varepsilon_-)=-\frac{2}{e^2}\int _{\mathscr{I}^+}\,\partial_{\bar{z}}\,\varepsilon_+(z,\bar{z})F_{uz} + \int _{\mathscr{I}^+} \,\varepsilon_+(z,\bar{z}) \,\gamma_{z\bar{z}}\,j_u \\
Q_-(\varepsilon_-)=-\frac{2}{e^2}\int _{\mathscr{I}^+}\,\partial_{\bar{z}}\,\varepsilon_-(z,\bar{z})G_{vz} + \int _{\mathscr{I}^+} \,\varepsilon_-(z,\bar{z}) \,\gamma_{z\bar{z}}\,j_v.
\end{align}
More explicit expressions for $Q_\pm$ will be derived in the appendix.
For Minkowski spacetime the matching is
\begin{equation}\label{matching}
Q_+(\varepsilon)=Q_-(\varepsilon), 
\end{equation} 
i.e. $\varepsilon_+=\varepsilon_-$. Note that the charge $Q_+$ is expressed in terms of out operators.\\

According to the philosophy of asymptotic quantization, it is assumed that Fourier components of field operators near the null infinity $\Scri$ create true eigenstates of the theory; in this case the incoming and outgoing scattering states of QED. With the non-trivial asymptotic dynamics in mind, this assumption has to be reviewed. In fact, below we will argue that the states created by field operators on $\Scri$ \textit{differ} from the true, dressed scattering states. As will be shown, it is precisely this difference that gives the connection between the decoupling of soft modes as in \ref{fulldecoupling} and the invariance under LGT. In order to highlight this difference, we will label the ``bare''  operators on $\Scri$  and states created by these with an additional $0$. This will be discussed in more detail in \ref{ScriExpansion}, we introduce the unusual notation here to avoid confusion in the rest of the paper.

The identity
\begin{equation}
\bra{p^0_{out}}Q_+(\varepsilon)-Q_-(\varepsilon)\ket{p'^0_{in}}=0
\end{equation}
implies (schematically)
\begin{align}
\int _{\mathscr{I}^+}\,\partial_{\bar{z}}\,\varepsilon\,\bra{p^0_{out}}\left(F_{uz}-G_{vz}\right)\ket{p'^0_{in}}= \nonumber \\
\frac{e^2}{2}\left(\gamma_{z(p)} \, \varepsilon(z(p))-\gamma_{z(p')} \, \varepsilon(z(p'))\right) \braket{p^0_{out}}{p'^0_{in}},
\end{align}
where we used the fact that $\ket{p^0_{in/out}}$ are eigenstates of $\int\, du\,j_u(u,z,\bar{z})$ and $\int\, dv\,j_v(u,z,\bar{z})$ respectively.\\

This is still not the end of the story since the operator $F_{uz}=F_{uz}(a^0_{out})$ is expressed in terms of (bare) out creation/annihilation operators. This is easy to amend using the identity 
\begin{equation}\label{inout}
a^0_{out}=\tilde{S}^\dagger\, a^0_{in} \, \tilde{S}.
\end{equation}
Since we are dealing with bare operators, here $\tilde{S}$ is the IR diverging $S$-matrix.

With that we finally obtain the scattering amplitude as originally written in \cite{strominger1} (but in a more extended notation):
\begin{align}\label{finally}
\int _{\mathscr{I}^+}\,\partial_{\bar{z}}\,\varepsilon\,\bra{p^0_{in}}\left( F_{uz}(a^0_{in})\,\tilde{S}-\tilde{S}\,G_{vz}(a^0_{in})\right)\ket{p'^0_{in}} = \nonumber \\
\frac{e^2}{2}\left(\gamma_{z(p)} \, \varepsilon(z(p))-\gamma_{z(p')} \, \varepsilon(z(p'))\right) \bra{p^0_{in}}\tilde{S}\ket{p'^0_{in}}.
\end{align}
Now everything is written in a single Fock space in terms of operators whose action is explicitly known. By using $\varepsilon(z)=\frac{1}{z-\zeta}$ one fishes out a single soft photon of a particular direction and polarization - then equation (\ref{finally}) reduces to the leading order soft photon theorem.

\subsection{Matching and symmetries}
In principle, the matching procedure can work and be useful even if there is no underlying symmetry - the identification of how one and the same operator acts on both in and out states can encode a lot of nontrivial information about a theory. However, in the above case the matching does correspond to a symmetry. The reason is simply the identity (\ref{inout}) which implies:
\begin{equation}
 \tilde{S}^\dagger\,Q_-(\varepsilon)\tilde{S}=Q_+(\varepsilon).
\end{equation}
Then, using the matching condition \ref{matching}, we find:
\begin{equation}
 \tilde{S}^\dagger\,Q_-(\varepsilon)\tilde{S}=Q_-(\varepsilon) \Longleftrightarrow [Q_-(\varepsilon),\tilde{S}\,]=0.
\end{equation}
In fact, the matching can be immediately derived from the soft symmetry as stated in (\ref{soft}).
The same situation is present for massive Fermions, where the construction of the asymptotic Fock spaces and the charges of LGTs is more complicated, see \cite{strominger2, strominger3} for a detailed treatment.

We want to stress again that in principle there can exist a matching that does not correspond to a symmetry but still gives rise to  interesting identities encoding the explicit breaking of the would be symmetry. For example, if we consider a process where an initial state gives rise to the formation of a black hole that subsequently evaporates completely, we can think of LGT charges defined independently on the in state and on the out state. Here the matching between the charges encodes the existence of the intermediate black hole resonance. If this matching is nontrivial, it will imply that the $S$ matrix controlling the whole process of creation and evaporation is not commuting with LGT or in other words that those LGT are explicitly broken for the $S$ matrix accounting for the evaporation. The scenario of a nontrivial matching that encodes information about the black hole was recently advocated in\cite{Hawking} and \cite{HPS}. We will address this question in more detail in a future publication.

\section{The role of large gauge transformations}\label{ScriExpansion}
We have seen that for the correctly dressed $S$ matrix the soft theorem is simply the statement that soft photons decouple from scattering. However, for the undressed dynamics the soft photon theorem seems less trivial and is in fact equivalent to the statement that LGT are symmetries of the scattering. How are the two sides of the story reconciled? \\

As announced above, in order to answer this question we need to revise the derivation of $Q_\pm(\varepsilon)$, in other words the expansion of fields next to null infinity $\Scri$. In the standard treatment, see \cite{Ashtekar1, Frolov}, it is assumed that fields evolve like free fields close to $\Scri$ and that their Fourier coefficients create true stationary scattering states. The existence of a nontrivial asymptotic dynamics should lead to a reexamination of this assumption. In order to do that let us first introduce some notation:\\

We denote the Fourier coefficients of the fields at a fixed spacial slice (say at t=0) by $a$ for the photon field and by $b, \, c$ for matter particles and antiparticles. They create the descriptor states of the ``free" Fock space by acting on the perturbative vacuum $\ket{0}$ . The dressed M\o ller operators are
\begin{equation}
\Omega(t) = \ee^{\iu\, H\, t}\,U_{as}(t) = \ee^{\iu\, H \, t}\ee^{-\iu\,H_0\,t}\ee^{R(t)} \quad ,\quad \Omega_{\pm}=\lim_{t\rightarrow \pm \infty} \Omega(t),
\end{equation}
with $R(t)$ as defined in (\ref{R(t)}). They accomplish:
\begin{equation}
\Omega_{\pm}\, a \, \Omega_{\pm}^\dagger = a_{in/out} \quad  , \quad \Omega_{\pm}\ket{0}=\ket{0}_{in}=\ket{0}_{out} 
\end{equation}
and the same for matter operators (we assume that the in and out vacua coincide for space-times without black holes). Obviously, this implies that for a descriptor state, say $
\ket{p}=b^\dagger(p)\ket{0}$ we get the corresponding stationary scattering state by applying M\o ller operators:
\begin{equation}
b^\dagger_{out/in}(p)\ket{0}_{out/in} \,=\,\ket{p_{out/in}}\,=\,\Omega_{\pm} \ket{p}.
\end{equation}
The key question now is the following: {\it are the Fourier coefficients of field operators close to null infinity $\Scri^-$ the creation/annihilation operators of true incoming particles, in particular $a_{in}$ for the photon field}? We shall argue that the Fourier coefficients of the photon field at $\Scri^-$ are {\it not} the full creation/annihilation operators of {\it true} incoming particles but rather the operators obtained assuming {\it trivial asymptotic dynamics}, let us call them $a_{in}^0$. In formulas the operators $a_{in}^0(k)$ are formally defined through
\begin{equation}
a^0_{in}(k)=\Omega^0_- \, a(k) \, \left(\Omega^0_-\right)^\dagger   \quad, \quad  \text{where } \Omega^0(t):=\ee^{\iu\, H \, t}\ee^{-\iu\,H_0\,t} .
\end{equation} 
Note that by the definition of $R(t)$ the (leading order) difference between $a_{in}(k)$ and $a_{in}^0(k)$ appears only in the $k\rightarrow 0$ limit and also that for the matter density
\begin{equation}
\rho^0_{in}(p)=\rho_{in}(p)
\end{equation} 
because $R(t)$ commutes with $\rho$.
In fact we have
\begin{align*}
&\lim_{k\rightarrow0}\,a_{in}(k)  =\lim_{t\rightarrow -\infty}\Omega^0(t) \lim_{k\rightarrow0}\ee^{R(t)} \, a(k) \, \ee^{R(t)^\dagger} \left(\Omega^0(t)\right)^\dagger = \nonumber \\
&\lim_{t\rightarrow -\infty}\Omega^0(t)  \,  \lim_{k\rightarrow0} \frac{e}{(2\pi)^3}   \left( a(k) - \int      \,  \frac{d^3p}{p\cdot k}    \left(\,p\,-\iu \, k\, J(p) \right) \rho(p)  \right)    \left(\Omega^0(t)\right)^\dagger  = \nonumber \\
&\lim_{k\rightarrow 0} \left( a^0_{in}(k) -  \int      \,  \frac{d^3p}{p\cdot k}    \left(\,p\,-\iu \, k\, J(p) \right) \rho_{in}(p)  \right) ,
\end{align*} 
so in short:
\begin{equation}\label{ain}
\lim_{k\rightarrow0}a_{in}(k)   =\lim_{k\rightarrow 0} \left( a^0_{in}(k)-  \int      \,  \frac{d^3p}{p\cdot k}    \left(\,p\,-\iu \, k\, J(p) \right) \rho_{in}(p)  \right).
\end{equation}
All indices were dropped in the above equations for a shorter notation.    \\

With the former relations at hand the role of LGT is surprisingly simple and clear and can be summarized by saying that even for the dressed dynamics the invariance under LGT's is still the soft photon theorem. 

Let us demonstrate how the proof works. First notice the following trivial identity:
\begin{equation}\label{symmetries}
[S,Q]=0 \Longleftrightarrow [\tilde{S},Q_{in}] =0
\end{equation}
with $\tilde{S}$ and $Q_{in}$ as in (\ref{Stilde}) and (\ref{Qin}). The form of the generators of LGTs is:
\begin{equation}
 Q_-(\varepsilon)=Q_{soft}\left(a^0_{in}\right)+Q_{hard}\left(\rho_{in} \right),
\end{equation}
where in our notation we suppressed the dependence on $\varepsilon$ on the right hand side. For the explicit form see (\ref{LGTcharges}).
Now from (\ref{ain}) and the linearity of $Q_{soft}$ in $a$ follows:
\begin{align}
 Q_{soft}(a_{in})=Q_{soft}(a_{in}^0)+Q_{hard}(\rho_{in}) = Q_-(\varepsilon).
\end{align}
See also section \ref{explanation}.
As the notation suggests, $Q_{soft}$ consists purely out of soft photons, so it commutes with the dressed $S$ matrix:
\begin{equation}
 [S_{KF}, Q_{soft}(a)]=0.
\end{equation}
Therefore relation (\ref{symmetries}) immediately implies that LGT's are conserved, i.e. 
$$[\tilde{S},Q_-(\varepsilon)]=0$$ 
(which in turn implies the antipodal matching). What we observe is that the conserved charge associated with the soft photon theorem, that for the undressed $S$ matrix is given by $Q_{soft}(a_{in}^0)+Q_{hard}(\rho_{in})$, becomes simply $Q_{soft}(a_{in})$ in the context of non-trivial asymptotic dynamics.\\

The physical meaning of the operators $a^0_{in}(k)$ can be understood by computing their action on a true incoming particle, say $b^\dagger(k)_{in}\ket{0}_{in}$, where $b^\dagger(k)$ creates an electron. Since by definition $a_{in}(k)\,b^\dagger_{in}\,(p)\ket{0}_{in}=0$ we can use (\ref{ain})  to find:
\begin{equation}\label{coherent}
\lim_{k^0\rightarrow 0} a^0_{in}(k)\,b^\dagger_{in}\,(p)\ket{0}_{in}=\, \lim_{k^0\rightarrow 0}\frac{1}{p\cdot k}\,\left( \,p -\iu \,k\, \,J(p) \, \right) \, \,b^\dagger_{in}\,(p)\ket{0}_{in}.
\end{equation}
which is telling us that \textit{true scattering states of QED} are coherent states (but only with respect to the ``free asymptotic'' photon operators $a^0_{in}(k)$) with infinitely many soft quanta \cite{Chung, Ashtekar1} .\\

Note that the operators of {\it true} incoming photons, $a_{in}$, in fact do {\it annihilate} the scattering state $\,b^\dagger_{in}\,(k)\ket{0}_{in}$ by definition. In summary, we always need to keep in mind that ``true" incoming photons states do not correspond to the operators $a^0_{in}$ but rather to the $a_{in}$. While working with the dressed $S$ matrix we don't need to use the null infinity operators  $a^0_{in}$; they can be used as a bridge between soft decoupling and LGT.\\

The dressing cloud reflects the electric field of the charged particle. This dressing is coherent and its {\it "constituents"} are the free asymptotic photon operators and not the true asymptotic photons of the infrared finite $S$ matrix. 

\section{The role of null infinity and normalizable states}\label{WhyScri}
Let us imagine that we ignore the existence of nontrivial asymptotic dynamics and consequently the difference between the soft operators  $a^0_{in}$ and $a_{in}$ discussed in the previous section. Incidentally, this point of view is the one usually pursued in the literature. Since the polarization is not important for this discussion, we shall not write out the index $\mu$ and all numerical factors will be ignored. 

The commutation relations of creation/annihilation operators on a spacial slice in spherical coordinates are:
\begin{equation}
\left[a(\omega,\vec{e}_{k}) \, ,\,a^\dagger(\omega',\vec{e}_{k'}) \right] = \frac{1}{\omega}\,\delta(\omega - \omega')\,\delta^2(\vec{e}_k-\vec{e}_{k'})
\end{equation}
where $\vec{e}_k$ is a unit vector in $\R^3$ and we use the notation $a(\omega,\vec{e}_k)=a(\vec{k}=\omega \cdot\vec{e}_k)$ with $\delta^2(\vec{e}_k-\vec{e}_{k'})$ being the delta function on $S^2$. A normalized 1-photon state in the free Fock space can be written as:
\begin{equation}
\ket{f}:=\int d\omega\,d^2\vec{e}_{k} \,f(\omega,\vec{e_k}) \,a^\dagger(\omega,\vec{e}_{k}) \ket{0}
\end{equation}
with 
\begin{equation}
1=\braket{f}{f}=\int d\omega\,d^2\vec{e}_{k}\,  \frac{\abs{f}^2}{\omega}.
\end{equation}
Let us focus on the frequencies and consider states of the type $f(\omega,\vec{e}_k)=f(\omega)g(\vec{e}_k)$ with $\int \, d^2 \vec{e}_k \,\abs{g}^2 =1$. The normalization condition for them becomes
\begin{equation}\label{norm}
1=\int d\omega\  \frac{\abs{f(\omega)}^2}{\omega}.
\end{equation}
We see that $f(w)$ must vanish sufficiently fast as $\omega\rightarrow 0$ for all normalizable states. 
The issue of normalizability becomes more important once the discussion is taken to $\Scri$ without first developing an IR-finite scattering theory. This can be reformulated by saying that the charge $Q^{soft}_{in}(\varepsilon)$ is non-zero in general. Indeed, the explicit expression for this charge contains the term $\omega \,\delta(\omega)\, a_{in}(\omega,\vec{e}_k)$ and clearly vanishes on all normalizable states, which satisfy (\ref{norm}). This situation creates a puzzle since classically the corresponding charge operator does not vanish in general, see \cite{Ashtekar1}. The solution to this puzzle lies in the realization that the classical non-vanishing soft factors are absorbed into the dressing of true scattering states through the non-trivial asymptotic dynamics. When working with the infrared finite $S$ matrix, non-normalizable states never enter. However, normalizable scattering states as created by superpositions of $b^\dagger_{in}$ appear as non-normalizable when they are written in terms of the $\Scri$ Fock space that is built with $a_{in}^0$. \\

This discussion explains why the matching of LGT was found by looking at $\Scri$ and not by the usual application of Noether's theorem to spacial slices. Indeed, e.g. on the spacial slice at $t=0$ (where the field operators create the descriptor Fock space) the soft symmetry of the dressed $S$ matrix is simply the {\it vacuum degeneracy} of the free Maxwell theory. In general, for every theory with quantum mechanically protected massless particles soft modes decouple from the infrared finite $S$ matrix and the vacuum degeneracy of the interacting theory coincides with the vacuum degeneracy of the free theory \footnote{Note that the former statement is also true for infrared finite theories with massless modes. A nice example is the Euler Heisenberg theory for photons obtained by integrating out fermion loops in standard QED. In this case, the asymptotic dynamics is trivial and the vacuum degeneracy is the one of the free theory.}.

\section{Summary and outlook}
 In theories with infrared divergences the corresponding asymptotic soft modes with $k=0$ are decoupled from the infrared finite $S$ matrix. These leads to a set of infrared symmetries parametrized by all possible polarizations and directions of these soft modes. Moreover, these symmetries can be mapped into LGT acting on the asymptotic states and commuting with the $S$ matrix. The soft components of these full fledged infrared symmetries are normally realized as spontaneously broken symmetries and account for vacuum degeneracy or hair, while the hard part accounts for the nontrivial asymptotic dynamics. \\

In the black hole case, the role of the soft modes and corresponding symmetries
depends on the particular approximation used. In  a full quantum treatment in which 
black hole formation (say in a two particle collision) and a subsequent evaporation process is treated as an $S$-matrix scattering process, say of the sort 
$2\rightarrow N$ \cite{2N}, the only relevant asymptotic symmetries are the ones 
of Minkowski.  On the other hand, if we try to describe some scattering process 
as an $S$-matrix process in the fixed classical background metric of an eternal black hole, the new soft (Goldstone-type) modes - corresponding to new symmetries that are spontaneously-broken by the black hole  geometry -  will enter the game. The candidates for such modes could be for example the $A$-modes identified in \cite{Amodes}.  \\


The decoupling of soft modes is exact in the approximation where self-interactions among them are ignored. The effect of  self interactions is to modify the coherent state nature of the asymptotic states defining the infrared finite $S$ matrix. Indeed, self interactions among the soft modes will generate cascades of soft modes where interaction can lead to new collective phenomena. The effects of soft self interaction for the dressing in theories like gravity could be possibly controlled by classicalization.
From a more formal point of view, a lot of work remains to be done, in particular a useful and IR safe diagrammatics that includes the dressing should be developed. This question has already been partially addressed, see e.g. \cite{Catana} and references therein. The results agree with the usual cross sections as calculated in the IR divergent theory, see also the discussion in appendix \ref{app2}.

\newpage
\begin{center}
\large
\textbf{Appendix}
\normalsize
\end{center}

\begin{appendix}
 \section{Generators of LGT}
 We will derive an explicit expression for the hard part of the charges \ref{LGTcharges} using the saddle point approximation and assuming free asymptotic dynamics. The soft part has been derived with great care in \cite{strominger1} and the hard part for massive QED in \cite{strominger3}. We were unable to find a similar calculation for massless matter and that explains why we present it here. We shall derive the charges for a scalar field $\phi$, the derivation for spinors is completely analogous. Factors of $2\pi$ will be ignored and for convenience we will only write out the particle part, the antiparticle part being included implicitly.

The first step is to assume free dynamics, in which case the Fourier components of $\phi$ close to $\Scri$ generate real scattering states. In other words, we write
\begin{equation}
 \phi(t,r,\vec{e}_x)=\int  \, p \,\ee^{\iu(t-r\vec{e}_x\cdot \vec{e}_p)p}\, b^\dagger(p,\vec{e}_p) \, dp\, d^2\vec{e}_p.
\end{equation}
Using advanced coordinates $u=t-r$ we obtain:
\begin{equation}
 \phi(u,r,\vec{e}_x)=\int  \, p \,\ee^{\,\iu\,u\,p}\,\ee^{\,\iu \,r\,p(1-\vec{e}_x\cdot \vec{e}_p)}\, b^\dagger(p,\vec{e}_p) \, dp\, d^2\vec{e}_p.
\end{equation}
For large $r$ we can use the saddle point approximation and perform the integration over the angles $d^2\vec{e}_p$. The relevant saddle is at $\vec{e}_p=\vec{e}_x$ and we obtain:
\begin{equation}
 \phi(u,r\gg 1, \vec{e}_x)\sim \frac{1}{r} \int  \, \,\ee^{\,\iu\,u\,p}\,\, b_{out}^\dagger(p,\vec{e}_x) \, dp \, ,
\end{equation} 
where we replace $b$ by $b_{out}$.
The hard part of the LGT on $\Scri^+$ is:
\begin{equation}\label{saddlepoint}
 Q_{hard}(\varepsilon)=\int \varepsilon(\vec{e}_x) \,j_u(u,\vec{e}_x) \, du \, d^2\vec{e}_x \, ,
\end{equation}
see equations 2.10 and 3.2 in \cite{strominger1}. Plugging (\ref{saddlepoint}) and a similar expression for the conjugated momentum $\pi(u,r\gg 1, \vec{e}_x)$ (including the antimatter part) into the expression for $j_u$ and integrating over $u$ one finds:
\begin{equation}
 Q_{hard}(\varepsilon)= \int \, d^3p \; \varepsilon(\vec{e}_p)\, \rho_{out}(\vec{p}),
\end{equation}
from which immediately follows
\begin{equation}
 Q_{hard}(\varepsilon)\, b^\dagger_{out}(p) \ket{0}_{out}=\varepsilon(\vec{e}_p)\, b^\dagger_{out}(p) \ket{0}_{out}.
\end{equation}
A similar procedure can be done at  $\Scri^-$ for incoming states. This is nothing but equation 7.2 from \cite{strominger1} where it was derived by semiclassical methods.

\section{A comparison to the usual treatment of IR divergences}\label{app2}

After the former discussion, the reader might wonder why a reexamination of such an old issue as the infrared structure of  QED is necessary at all. 
The problem of IR divergences in QED was reasonably well understood after the seminal work of Bloch and Nordsiek in 1937 \cite{Bloch}, and of Yennie, Frautschi and Suura in 1961 \cite{YFS}. In order to highlight the difference between the dressing approach and the usual strategy for dealing with infrared divergences, we will briefly review the key ingredients of the old approach in this appendix. The present discussion closely follows \cite{YFS}.

\subsection{Factorization and resummation}
Let us start by considering the simplest amplitude in QED, an initial one electron state $|p\rangle$ going over into a final state $|p'\rangle$, call it $M(p,p')$. This amplitude is a sum of diagrams that we can characterize in terms of the number of internal virtual photon lines. Thus, generically we have $M(p,p')= \sum_nM_n(p,p')$. Here $n$ is the number of virtual photons and $M_n$ is the sum of all diagrams ( contributing to the process ) with $n$ internal virtual photon lines. These diagrams are infrared divergent. A careful analysis of the infrared divergences allows to obtain the infrared divergent part of $M_{n+1}$ recursively in terms of that of $M_{n}$. After imposing symmetrization over the involved virtual photons, an infrared resummation can be performed leading to a factorized form of the amplitude:

\begin{equation}
M(p,p') = e^{\alpha B(p,p')}  \tilde M(p,p'),
\end{equation}
with $\tilde M$ being infrared finite. The infrared divergent part $B$ can be written in terms of an infrared regulator, e.g. non-zero photon mass $m$. The behavior of $B$ for this simplest process with just one electron in the in and out states and for very large $p$ and $p'$  is (to leading order in $m$): 
\begin{equation}
B\approx- \frac{1}{2\pi}\left( \ln\frac{2p\cdot p'}{m_e^2} \left(\ln\frac{m_e^2}{m^2} -1/2 +1/2 \ln\frac{2p.p'}{m_e^2}\right) - \ln\frac{m_e^2}{m^2}\right).
\end{equation}One can immediately observe that the regularized infrared divergent part  goes like
 \begin{equation}
 B\sim - \left(\ln\frac{2p\cdot p'}{m_e^2} -1\right)\ln \frac{m_e^2}{m^2}  \, \xrightarrow{m\rightarrow 0} \, -\infty 
 \end{equation}
 and therefore leads to zero amplitudes in the limit of zero mass photon. The fact that the infrared divergent part factorizes after resummation, allows to define a sort of infrared ``renormalization'' by simply defining the physical amplitude to correspond to the finite part. This common sense recipe is essentially correct, but it needs a physical justification.
 
 \subsection{Real photon emission and unitarity}
 Let us now consider the same amplitude as above with an additional real photon of momentum $k$ in the in or out state. This amplitude is infrared divergent in the $k\rightarrow0$ limit. Contrary to the divergences associated to virtual photon lines, this is a divergence for an a priori perfectly well defined amplitude where the in and out states are fixed. If we take this infrared divergence seriously (i.e. do not speak about imperfect detectors), we could conclude that the corresponding amplitude violates unitarity for soft emission. In order to understand this apparent violation of unitarity, one can compute the physically meaningful quantity 
 \begin{equation}
 \int d^3k \frac{1}{\omega(k)} \delta(\epsilon-k)\abs{ \bra{p, E} S_D \ket{p',E',k}}^2
 \end{equation}
 where $\epsilon = E'-E$. This is the probability to emit a real photon with total energy $\epsilon$. Using the soft photon theorem and the integral representation of the delta function, the infrared divergent part of the previous integral can be written as 
 \begin{equation}
  \tilde B = \int_{0}^{\epsilon} d^3k \frac{1}{\omega(k)} (\frac{p'}{k.p'} - \frac{p}{k.p})^2
 \end{equation}
 Introducing an infrared cutoff $K$ in the integral one obtains
 \begin{equation}
 \left(\ln\frac{2p.p'}{m_e^2} -1\right) \ln\frac{\epsilon^2}{K^2}
 \end{equation}
 In this expression there are two types of problems: There is an infinity when the IR cutoff $K$ is sent to zero and there is an UV problem even for finite $K$ due to the logarithmic growth of $\ln\frac{2p.p'}{m_e^2}$ with the energy. This UV growth is in principle a real problem even for a theory with a natural IR cutoff, like a finite photon mass. So what to do with this unitarity problem?\\
 
 The standard solution comes from a  resummation of \textit{different} processes (contrary to the resummation of different diagrams for the same process, as mentioned above). Indeed, let us consider amplitudes with $n$
 real soft photons in the external lines. For each value of $n$ the corresponding differential cross section is defined by integrating over the phase space of the emitted photons and symmetrizing. An interesting fact is that for these differential cross section we can, as we did for the amplitudes before, perform a resummation over $n$. After this resummation is done at the level of the cross section, one finds that the infrared divergent part again factorizes. The philosophy of imperfect detectors suggests that only inclusive cross sections are relevant, so the exponential factors for a particular process coming from virtual and real photons must be combined into the form
 \begin{equation}
 e^{\alpha (B + \tilde B)}
 \end{equation}
 with $B$ and $\tilde B$ exactly canceling the infrared divergent parts. So we see that thanks to the resummation (i.e. imperfect detectors),the divergent parts can be exponentiated in such a way that the virtual and real components cancel each other. \\
 
 In the above solution \textit{unitarity is lost for individual processes} and is re-obtained only for inclusive cross sections. Complementary to that, in the dressing scenario, as advocated in the present paper, every process is IR safe and inclusive cross sections become unnecessary since soft particles decouple from scattering.

 \subsection{The dressing interpretation of the resummation over real emission}
 We now want to comment on the connection between the standard resummation approach to IR diverges and the dressing approach.
 The fact that a resummation (at the level of the differential cross section) over the number of real emitted photons can be performed, allows to reinterpret the IR divergences by defining a new charged asymptotic state as a {\it dressed state} of real soft emitted photons. One can hope that with a correct dressing the unitarity problem for fixed processes disappears since the dressing should already account for resummation. The strategy of Kulish and Faddeev \cite{FK} in 1970, which is also our strategy in the present paper, was to find the correct dressing from asymptotic dynamics and not from the IR divergences in the Dyson $S$ matrix. Nevertheless, these issues are clearly intertwined. Loosely speaking, one can think of  the Dyson $S$ matrix containing the infrared part $e^B$, while the part $e^{\tilde B}$ accounting for the resummation over real emitted photons is absorbed into the coherent state definition of the asymptotic states. The fact that the infrared divergent parts of $B$ and $\tilde B$ cancel explains why the matrix $S_{KF}$ is infrared finite. \\
 
 It is the universality of infrared physics that allows for the amazing simplicity of the resummation. This universality appeared in different forms throughout the present paper. It was seen in the soft theorem, the definition of asymptotic dynamics, the decoupling of soft modes and finally the matching and the invariance under LGT. In the present work we have tried to illuminate the connections between (and the equivalence of) these seemingly different manifestations of infrared physics.
 
 The former discussion sheds also light on the roots of LGT and infrared symmetries of QED. The resummation and exponentiation of the infrared divergences associated with real emitted soft photons promotes the soft photon theorem ( and its subleading component ) into the formal {\it generator} of a transformation. On the other hand the infrared cancellation between real and virtual divergences makes this transformation a symmetry of the $S$ matrix. 
 
%
 
 \end{appendix}

\section*{Acknowledgements}
We want to thank Gia Dvali for the many fruitful discussions and ongoing collaboration. We also thank Kepa Sousa, Daniel Flassig, Claudio Bunster, Raoul Letschka and Artem Averin for valuable discussions and comments. 
The work of C.G. was supported in part by Humboldt Foundation and by Grants: FPA 2009-07908, CPAN (CSD2007-00042) and by the ERC Advanced Grant 339169 "Selfcompletion'' .
The work of M.P. was supported by the ERC Advanced Grant 339169 "Selfcompletion''.

\end{document}